\documentclass{aa}

\usepackage{amssymb,amsmath}
\usepackage{graphicx}
\usepackage{natbib}
\usepackage{bm}
\usepackage{booktabs}
\usepackage{url}
\usepackage{xcolor}

\bibpunct{(}{)}{;}{a}{}{,} 

\def\fm{fm$^{-3}$}

\def\beq{\begin{equation}}
\def\eeq{\end{equation}}
\def\beqn{\begin{eqnarray}}
\def\eeqn{\end{eqnarray}}

\begin{document}
\title{Crystallization of the inner crust of a neutron star and the influence of shell effects
}
\titlerunning{Title}
\authorrunning{Carreau et al.}

\author{T. Carreau\inst{1},  
F. Gulminelli\inst{1},
N. Chamel\inst{2}, 
A.~F. Fantina\inst{3,2},
J.~M. Pearson\inst{4}}

\institute{LPC (CNRS/ENSICAEN/Universit\'e de Caen Normandie), UMR6534, 14050 Caen C\'edex, France \\
  \email{carreau@lpccaen.in2p3.fr}
\and Institut d'Astronomie et d'Astrophysique, CP-226, 
Universit\'e Libre de Bruxelles, 1050 Brussels, Belgium \\
\and Grand Acc\'el\'erateur National d'Ions Lourds (GANIL), CEA/DRF -
 CNRS/IN2P3, Boulevard Henri Becquerel, 14076 Caen, France \\
\and D\'ept. de Physique, Universit\'e de Montr\'eal, Montr\'eal, Qu\'ebec H3C 3J7 Canada
}
   
\date{Received xxx Accepted xxx}

\abstract{
In the cooling process of a non-accreting neutron star, the composition and
properties of the crust are thought to be fixed at the finite temperature where
nuclear reactions fall out of equilibrium. A lower estimation for this
temperature is given by the crystallization temperature, which can be as high
as {$\approx 7\times 10^9$ K} in the inner crust, potentially leading to sizeable differences with respect to the simplifying {cold-}catalyzed matter hypothesis.
}
{
We extend the recent work by \citet{fantina2019} on the outer crust, to the study of the crystallization of the inner crust and the associated composition in the one-component plasma approximation.
} 
{
The finite temperature variational equations for non-uniform matter in both the liquid and the solid phases are solved using a compressible liquid-drop approach with parameters optimized on four different microscopic models which cover the present uncertainties in nuclear modeling. 
}
{
We consider separately the effect of the different nuclear ingredients with their associated uncertainties, namely the nuclear equation of state, the surface properties in the presence of a uniform gas of dripped neutrons, and the proton shell effects arising from the ion single-particle structure. {Our results suggest that the highest source of model dependence comes from the smooth part of the nuclear functional.}
}
{
We show that shell effects play an important role at the lowest densities close to the outer crust, but the most important physical ingredient to be settled for a quantitative prediction of the inner crust properties is the surface tension at extreme isospin values.
} 


\maketitle

\section{Introduction}
\label{sect:introd}

The essential input determining the composition of the outer crust of a cold non-accreting neutron star (NS) under the cold-catalyzed matter hypothesis is given by 
the masses of the atomic nuclei which are confined to the crystalline ion sites. These masses are experimentally measured to a high level of accuracy, meaning that the
{properties}
of the outermost layers of the crust are {precisely}
known \citep{Blaschke2018}. 
However, for the regions deeper in the star, matter becomes so neutron rich that experimental data are unavailable,
and model dependence arises. 
{The uncertainty due to modeling becomes a critical aspect} 
in the inner crust, {extending from} $\approx 300$~m {below the surface} 
down to 
to about 1 km in depth; {in this region} neutrons drip off nuclei forming a gas, 
a situation which cannot be reproduced in the laboratory. 

In the {inner crust} regime, the properties of matter depend on the energetics of both neutron matter and extremely exotic neutron-rich nuclei, which can only be accessed by nuclear modeling. In turn, the mass of dripline nuclei as calculated by nuclear models depend on the bulk properties of asymmetric matter\footnote{The asymmetry we refer to in this paper is in terms of neutron/proton ratio.} as expressed by the so-called nuclear equation of state (EoS), but also on the details of nuclear structure. {These} 
include surface properties arising from the finite size of the nucleus, as well as shell corrections arising from the underlying single-particle structure of the nuclei. 

In the recent years, a huge progress was made in constraining the properties of the nuclear EoS from astrophysical observations, ab-initio calculations and nuclear experiments, see \citet{Burgio2018} for a recent review. 
It is therefore interesting to study how much these constraints impact the uncertainties in the predictions of the composition of the inner crust. 

An extra complication arises from the fact that the crust {of a NS is unlikely to be in full thermodynamic equilibrium at zero temperature. }
In reality, NSs are born hot, and if {their} core cools down sufficiently rapidly,  the composition might be frozen at a finite temperature, see e.g. \citet{goriely2011}. 
Deviations from the ground-state composition in the cooled crust around the neutron-drip density were already considered in \citet{bisno1979}, but simple extrapolations of semi-empirical mass formulas were used at that time. 
The value of the freeze-out temperature is difficult to evaluate, but a lower limit is given by the crystallization temperature, since we can expect that nuclear reactions will be fully inhibited in a Coulomb crystal.

In \citet{fantina2019},  the crystallization of the outer crust of a non-accreting unmagnetized NS in the one-component plasma (OCP) approximation {has been} recently studied, using the microscopic HFB-24 nuclear {mass model}~\citep{goriely2013}. 
The underlying functional {BSk24 has been} also recently used to determine the ground-state composition and the equation of state in all regions of a non-accreting NS by \citet{pearson2018}. 
In the present paper, we extend the work of \citet{pearson2018} and \citet{fantina2019}, by calculating the crystallization temperature and the associated composition in the inner crust in the OCP approximation. 
To this end, we solve the variational equations for non-uniform matter within the compressible liquid-drop {(CLD)} approach presented in \cite{Carreau2019}. This formalism is presented in Section \ref{sect:model}.
The use of the CLD approach will allow us to additionally address the question of the model dependence of the results. 
Indeed in this semi-classical approach  shell effects are added on top of the smooth EoS functional by using the Strutinsky Integral method, as explained in Section \ref{sect:shell}. This allows an independent variation of the bulk parameters, the surface parameters, and the shell effects, and a comparison of the relative importance of the three different microscopic ingredients. This will be {discussed} in 
Section \ref{sect:results}, by considering four different realistic microscopic functionals providing comparably good reproduction of {various experimental and theoretical nuclear data, }
taken from \citet{goriely2013} and \citet{pearson2018}. Conclusions are finally drawn in Section \ref{sect:conclus}.

\section{Model of the inner crust}
\label{sect:model}

{In the homogeneous matter limit,} an EoS model corresponds to a given free energy functional for bulk nuclear matter at proton (neutron) density $n_p (n_n)$ and temperature $T$, $f_b(n_B,\delta,T)$, with $n_B=n_p+n_n$, $\delta=(n_n-n_p)/n_B$ and $f_b$ the free energy per baryon.  We use the four different functionals of the BSk family taken from \citet{goriely2013}, 
namely BSk22, BSk24, BSk25, and BSk26. These models are chosen because they all
provide excellent fits to the {2016 Atomic Mass Evaluation (AME) \citep{ame2016}}, are compatible both with ab-initio and NS mass
constraints, and explore a relatively large domain in the symmetry energy
parameters (consistent with existing experimental constraints), which
constitute the most important part of the EoS uncertainty \citep{pearson2014, pearson2018}. 
Moreover, full mass tables obtained {by} complete
Hartree-Fock-Bogoliubov (HFB) calculations {have been published for these models\footnote{The mass tables are
available on the BRUSLIB online database \url{http://www.astro.ulb.ac.be/bruslib/}~\citep{xu2013}.}}.

Once the EoS model is defined, the equilibrium configuration of inhomogeneous dense matter in the inner crust in full thermodynamic {equilibrium}  
at temperature $T$ and baryon density $n_B$ is obtained following \citet{lattimer1991,gulrad2015}, {who extended} 
to finite temperature the variational formalism of ~\citet{bps,Douchin2001}.

Using the Lagrange multipliers technique, the free-energy density in a Wigner-Seitz cell of volume $V$ is minimized with the constraint of a given baryon density $n_B$. The auxiliary function to be minimized reads
\begin{equation}
\mathcal{F}
=\frac{F_{i}}{V}+\left (1-\frac{A}{n_0 V}\right )\mathcal{F}_g+\mathcal{F}_{e}-\mu n_B, \label{eq:auxiliary}
\end{equation}
where $\mathcal{F}_g=n_g f_b(n_g,1,T) {+ n_g m_nc^2}$ ($\mathcal{F}_{e}$) is the free energy density\footnote{We use capital letters for the energy per ion, i.e. $F$ is the ion
free energy, small letters for the energy per baryon, i.e. $f$ is the free
energy per baryon, and the notation $\mathcal{F}$ for the free energy
density.}
 of a pure uniform neutron (electron) gas at density $n_g$ ($n_e$),
and the bulk interaction between the nucleus and the neutron gas is treated in the excluded volume approximation. {As discussed by \citet{pearson2012}, minimizing $F$ at fixed baryon density is pratically equivalent to minimizing the Gibbs free energy $G$ at fixed pressure. } 
The term $F_i$ corresponds to the free energy of a fully ionized atom of mass number $A$, atomic number $Z$, and density $n_0$, including the Coulomb (nuclear) interaction with the electron (neutron) gas, and is given by (see Chap.~2 in \citet{hpy2007})
\begin{equation}
\label{eq:Fi}
F_i = \left (Zm_p+(A-Z)m_n\right )c^2 + F_{\rm nuc} + F_i^{\rm{id}} + F_i^{\rm int} \ ,
\end{equation}
where $m_{n(p)}$ is the neutron (proton) mass, $c$ being the speed of light, $F_{\rm nuc}$ {is} the free energy associated to the nuclear and electrostatic interactions among nucleons, $F_i^{\rm{id}}$ is the non-interacting (``ideal'') contribution to the ion free energy, and $F_i^{\rm int}$ accounts for Coulomb electron-ion and electron-electron interactions.

Depending on the phase, either solid or liquid, different expressions enter in
the ideal $F_i^{\rm{id}}$ and non-ideal $F_i^{\rm int}$ free energy terms,
{apart
from the finite-size contribution which is common to both phases.} 
The full expressions for the different terms can be found in \citet{fantina2019}.
The most important term determining the transition from the liquid to the solid phase is $F_i^{\rm{id}}$, specifically
 the zero-point quantum-vibration
 term in the solid phase, and the translational term in the liquid phase. 
Exchange and polarization corrections are found to have no effect in the density and temperature regime studied in the present paper and {are therefore} neglected. We also ignore the possible presence of free protons, which are expected to be negligible in 
the low temperatures and proton fractions characterizing the inner crust around the crystallization point.

The nuclear free energy $F_{\rm nuc}$ is the same in the liquid and solid phases, and it is 
calculated in the {CLD} approximation,

\begin{equation}
    F_{\rm nuc}^{\rm CLD} = Af_{b}(n_0,I,T)   + F_{\rm {surf+curv}} + F_{\rm Coul}, \label{eq:fnuc}
\end{equation}
where the bulk neutron-proton asymmetry is given by $I=1-2Z/A$.

Assuming spherical geometry, we write the Coulomb energy as
\begin{equation}
  F_{\rm Coul}= \frac{3}{5} \frac{e^2}{r_0} \frac{Z^2}{A^{1/3}},
\end{equation}
with $r_0 = (4\pi n_0/3)^{-1/3}$, $e$ the elementary charge, and the surface and curvature free energies as in~\citet{Newton2013}
\begin{eqnarray}
   \label{eq:Fsc}
    F_{\rm {surf+curv}} & = & 4\pi r_0^2\sigma_s A^{2/3} \nonumber \\
    & & + 8\pi r_0\sigma_s\frac{\sigma_{0,c}}{\sigma_0}
    \alpha\left(\beta-\frac{1-I}{2}\right)A^{1/3}, 
\end{eqnarray}
with $\alpha = 5.5$. The expression for the surface tension $\sigma_s$  is given in ~\citet{lattimer1991} and is suggested from Thomas-Fermi calculations in the Wigner-Seitz cell in the free neutron regime,
\begin{equation}
    \sigma_s = \sigma_0 
    \frac{2^{p+1} + b_s}{(Z/A)^{-p} + b_s + (1-Z/A)^{-p}}h(T). \label{eq:sigma}
\end{equation}
{In Eqs.~(\ref{eq:Fsc})-(\ref{eq:sigma}),} the surface and curvature parameters $\sigma_0$, $b_s$, $\sigma_{0,c}$, and $\beta$ 
govern the surface properties of nuclei at moderate asymmetries below neutron drip.
Following the same strategy as in \citet{Carreau2019}, they can be fixed  by fitting the $T=0$  limit of Eq.~(\ref{eq:fnuc}) 
{to a given mass table. For this study, we have built full}
 {fourth-order} Extended Thomas-Fermi (ETF) mass tables for each of the functionals BSk22, BSk24, BSk25, or BSk26{, and fitted the parameters of Eq.~(\ref{eq:sigma}) to the ETF results.\footnote{The ETF mass tables of the models  BSk22, BSk24, BSk25, {and} BSk26 are given as supplementary material to this work.}} 
 The function $h$ effectively accounts for the excitation of surface modes at finite temperature, $h(T > T_c) = 0$ and $h(T \leq T_c) = \left(1 - \left(\frac{T}{T_c}\right)^2\right)^2$, see Eq. (2.31) of ~\cite{lattimer1991}. 
Given 
that the critical temperature $T_c$ is of the order of {$T_c \approx 1.75\times
10^{11}$ K,}
and the crystallization temperature is lower than {$10^{10}$ K} in the inner crust, $h \approx 1$ and the excitation of surface modes {can thus be} neglected.

Finally, the $p$ parameter determines the behavior of the surface tension at high isospin above the neutron drip, 
and effectively accounts for the nuclear interaction between the nucleus and the surrounding neutron gas, which can be modeled as a surface effect. We fix it to reproduce 
the crust-core transition density for the four BSk functionals considered. Those {transition} points were obtained by~\cite{pearson2019} using the same method as in~\cite{ducoin2007}, with the transition taking place when homogeneous NS matter becomes unstable against finite-size fluctuations. The corresponding optimized surface and curvature parameters are displayed in Table~\ref{tab:surfpar} (see also Table 14 of~\cite{pearson2019}).

\begin{table}[ht]
   \caption{Surface and curvature parameters optimized to reproduce the 
      crust-core transition properties of functionals BSk22, BSk24, BSk25, and 
  BSk26.}
  \begin{tabular}{lccccc}
    \toprule
    & $p$ & $\sigma_0$ & $b_s$ & $\sigma_{0,c}$ & $\beta$ \\
    & & (MeV/fm$^2$) & & (MeV/fm) & \\
    \midrule
    BSk22 & 3 & 0.99785 & 23.073 & 0.10403 & 1.0717 \\
    BSk24 & 3 & 0.98636 & 36.227 & 0.09008 & 1.1631 \\
    BSk25 & 3 & 0.98964 & 47.221 & 0.08902 & 1.1483 \\
    BSk26 & 3 & 0.98797 & 34.346 & 0.10985 & 1.0609 \\ 
    \bottomrule
  \end{tabular}
\label{tab:surfpar}
\end{table}

The quality and flexibility of the parametrization Eq.~(\ref{eq:sigma}) was in particular verified by \citet{Newton2013}, showing
that the seminal crust composition of \citet{bbp} can be indeed reproduced with it.

Within the CLD approximation, at a given baryon density and temperature, the nuclear free energy $F_{\rm nuc}$ solely depends on the three parameters $(A, I, n_0)$, while the global free energy density additionally depends on the electron and free neutron densities, $n_e=n_p$ {and} $n_g$. 
 
The equilibrium configuration is thus obtained by independent variations of the auxiliary function {Eq.}~(\ref{eq:auxiliary})
with respect to the five variables $A,I,n_0,n_p,n_g$
using the baryon density constraint,
\begin{equation}
n_B=n_g+\frac{A}{V}\left (1-\frac{n_g}{n_0} \right ).
\end{equation}

This leads to the following system of coupled differential equations:

\begin{eqnarray}
    \frac{\partial (F_{i}/A)}{\partial A}
& = &0, \label{eq:eq1} \\
    \frac{2}{A}\left ( \frac{\partial F_{i}}{\partial I}
- \frac{n_p}{1-I}
 \frac{\partial F_{i}}{\partial n_p}
\right ) + \Delta m_{n,p}c^2
         &=& \mu_{e}^{\rm tot} , \label{eq:eq2} \\
    \frac{F_{i}}{A} + \frac{1-I}{A}\frac{\partial F_{i}}{\partial I}
&=& \mu - \frac{P_g}{n_0}, \label{eq:eq3} \\
    {n_0}^2\frac{\partial (F_{i}/A)}{\partial n_0}
&=& P_g, \label{eq:eq4}
\end{eqnarray}

\noindent where {$\mu_e^{\rm tot}$ is the electron chemical potential including the rest mass (see, e.g., Sect.~2.3.1 in \citet{hpy2007}), $\Delta m_{n,p}$ is the neutron-proton mass difference,}
 the gas pressure is given by $P_g=n_g\mu-\mathcal{F}_g$, 
and the baryon chemical potential $\mu=\mu_B^{\rm tot}-m_nc^2$ results:

\begin{equation}
    \mu = \frac{2 n_pn_0}{n_0(1-I) - 2 n_p}\frac{\partial (F_{i}/A)}{\partial n_g}
+ \frac{d\mathcal{F}_g}{dn_g} .
\end{equation}

In our parametrization, the in-medium modification of the surface energy arising from the external gas is governed
by a single parameter $p$ which does not depend on the external neutron density but only on the global asymmetry. 
Then $\partial F_{i}/\partial n_g=0$ and the baryon chemical potential
can be identified with the chemical potential of the gas $\mu_g\equiv d \mathcal{F}_g/ d n_g$.

At each value of the baryon density $n_B$ and temperature $T$,
the system of coupled differential equations Eqs.~(\ref{eq:eq1})-(\ref{eq:eq4}) is numerically solved as in 
~\cite{Carreau2019}. Specificaly, we solve the coupled equations  using the expressions of $F_i^{\rm id}$ and $F_i^{\rm int}$ of the liquid phase, yielding the optimal liquid composition $\bm{X_{\rm liq}} = (A_{\rm liq}, I_{\rm liq}, n_{0,{\rm liq}}, n_{g,{\rm liq}})$ and the associated free energy density $\mathcal{F}_{\rm liq}$. 
Then, for the same composition $\bm{X_{\rm liq}}$, we calculate the free energy density assuming a solid phase $\mathcal{F}_{\rm sol}$. The {lowest} 
temperature for which $\mathcal{F}_{\rm liq} \ge \mathcal{F}_{\rm sol}$ is identified as the crystallization temperature $T_{\rm m}$
corresponding to the baryon density under study.

\section{Inclusion of shell effects}
\label{sect:shell}

\begin{figure}[htbp]
\begin{center}
\includegraphics[width=\linewidth]{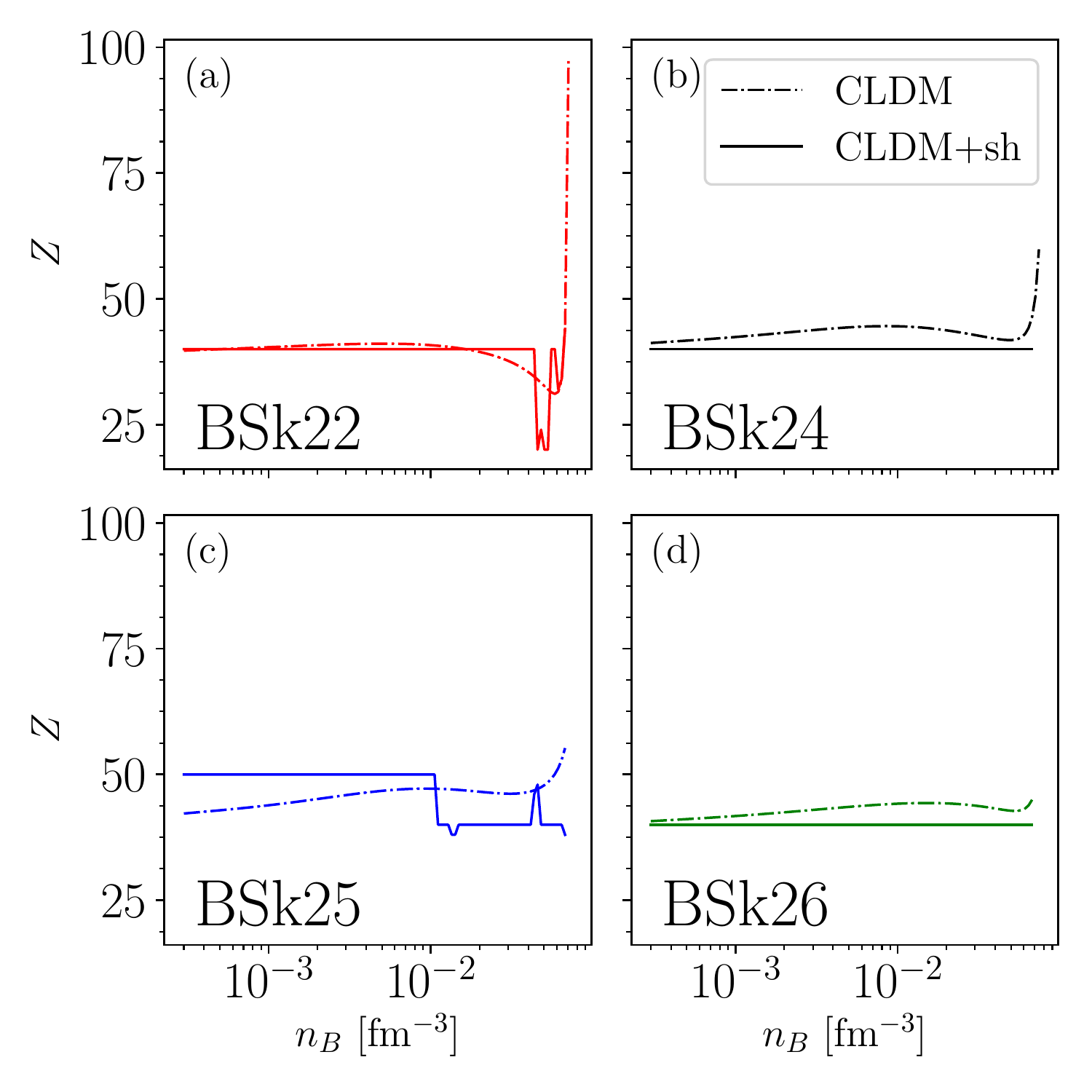}
\end{center}
\caption{Equilibrium value of proton number $Z$ as a function of baryon density in the inner crust at zero temperature, for the BSk22 (panel (a)), BSk24 (panel (b)), BSk25 (panel (c)), and BSk26 (panel (d)) {CLD models}, with (solid lines) and without (dot-dashed lines) shell effects. 
See text for details.}\label{fig:compo_T0}
\end{figure}

\begin{figure*}[htbp]
\begin{center}
\includegraphics[width=\linewidth]{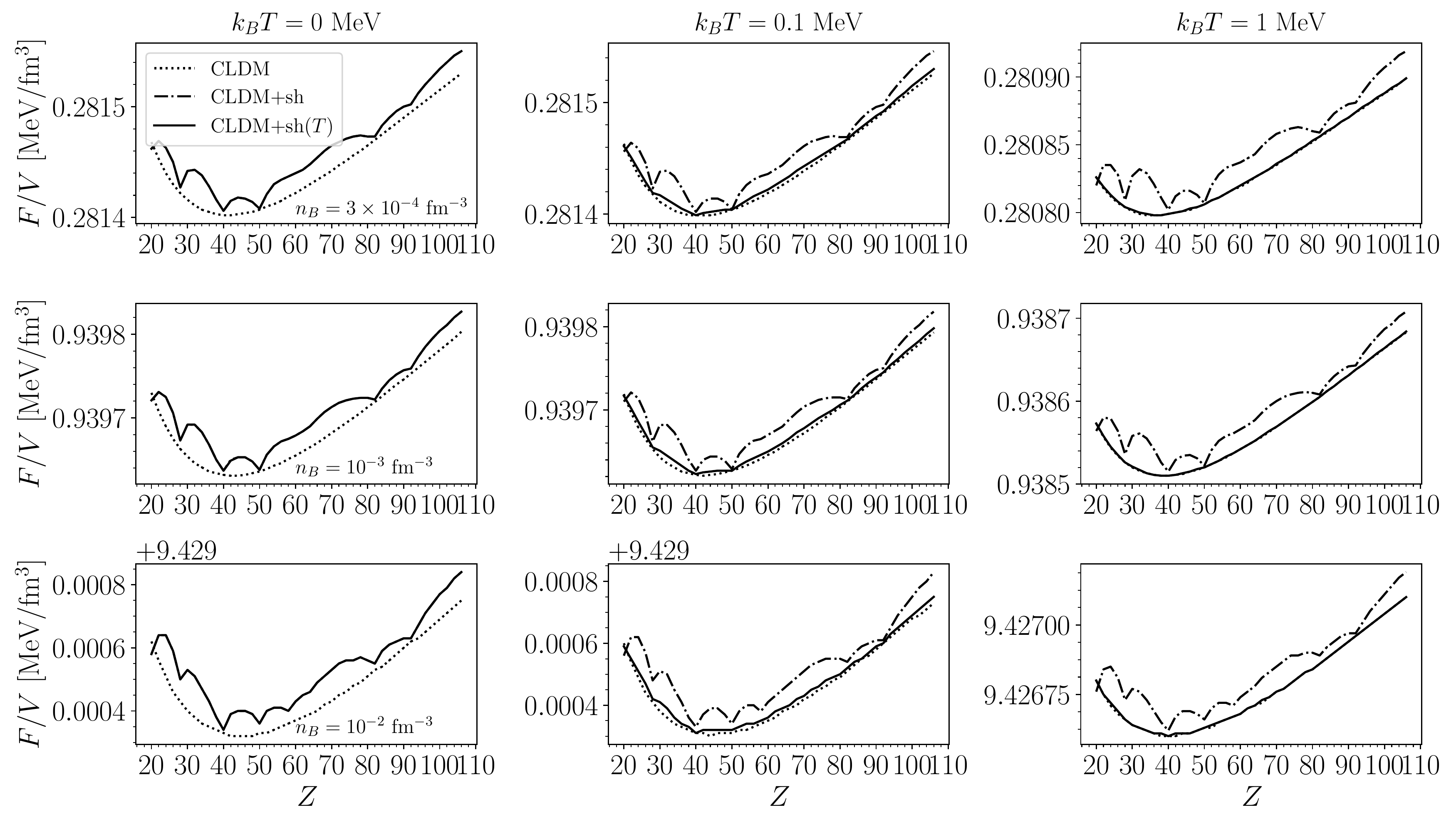}
\end{center}
\caption{Free energy density as a function of the proton number $Z$ for the
  BSk24 {CLD} model for different temperatures: $k_B T=0$ MeV (left panels), $k_B
  T=0.1$~MeV (central panels), and $k_B T=1$~MeV (right panels), and densities: $n_B = 3 \times 10^{-4}$~\fm~(top panels), $n_B = 10^{-3}$~\fm~(middle panels), and $n_B = 10^{-2}$~\fm~(bottom panels). In each panel, dash-dotted (solid) lines correspond to the calculations including temperature-independent (temperature-dependent) shell effects, and dotted
lines represent the results without shell effects. See text for details.}\label{fig:fws}
\end{figure*}

In the CLD approach, shell effects, which are known since the pioneering work of \citet{bps} to be essential to correctly evaluate the outer-crust composition in the $T=0$  limit, are lost.
In this same limit, the microscopic calculations of \citet{chamel2006},~\citet{chamel2007} have shown that the neutron-shell effects {become vanishingly small beyond the neutron-drip point}, 
but proton shell effects persist in the inner crust. 
Those shell corrections have been calculated using the Strutinsky Integral method in \citet{pearson2018} for the same functionals as used in this paper. 
We have therefore added these corrections\footnote{Updated tables containing the ETFSI energy thus the shell corrections for the BSk22, BSk24, BSk25, and BSk26 are given as supplementary material in \citet{pearson2019}.} on top of the CLD free energy, Eq.~(\ref{eq:fnuc}).
In this way, we can recover the expected appearance of the magic numbers in the composition for the inner crust, as shown in Fig.~\ref{fig:compo_T0}, where results for the composition at zero temperature for the CLD calculations (dot-dashed lines) are compared with those including Strutinsky shell corrections for the corresponding functional (solid lines). 
Indeed, the continuous smooth behavior of $Z$ in the inner crust obtained in the CLD approach disappears when shell effects are added.
In particular, remarkable stability at $Z=40$ ($Z=40$ and $Z=20$) is seen for the BSk24 and BSk26 (BSk22) model, in very good agreement with \citet{pearson2018} (see their Fig.~12).
A small difference only appears for the BSk25 model, reflecting the limitations of the CLD approach. For this functional, after the plateau at $Z=50$, also obtained with the ETF calculations of~\cite{pearson2018} including proton shell effects with the Strutinsky integral method (ETFSI) for the same functional, $Z$ drops to $Z=40$ in our case, while increasing values of $Z$ are observed in~\cite{pearson2018}. {However,~\cite{pearson2019} find a secondary minimum at Z=40, and the energy difference between the two minima is of the order of $10^{-3}$ MeV.} 

To analyze the effect of the shell corrections in the free energy, we plot in Fig.~\ref{fig:fws} the free energy density as a function of $Z$ for different densities in the inner crust and for three selected temperatures. 
We notice that, as expected, the pure CLD results are close to those including zero-temperature shell corrections only for closed-shell configurations, while remarkable differences exist for all other values of $Z$.
This confirms the importance of a proper account of the shell structure.
The increasing discrepancy at $k_B T=1$~MeV, with $k_B$ the Boltzmann
constant, is instead due to an overestimation of the shell corrections, which are expected to be wiped out at (high enough) finite temperature, as we discuss next.

\begin{figure}
\begin{center}
\includegraphics[width=\linewidth]{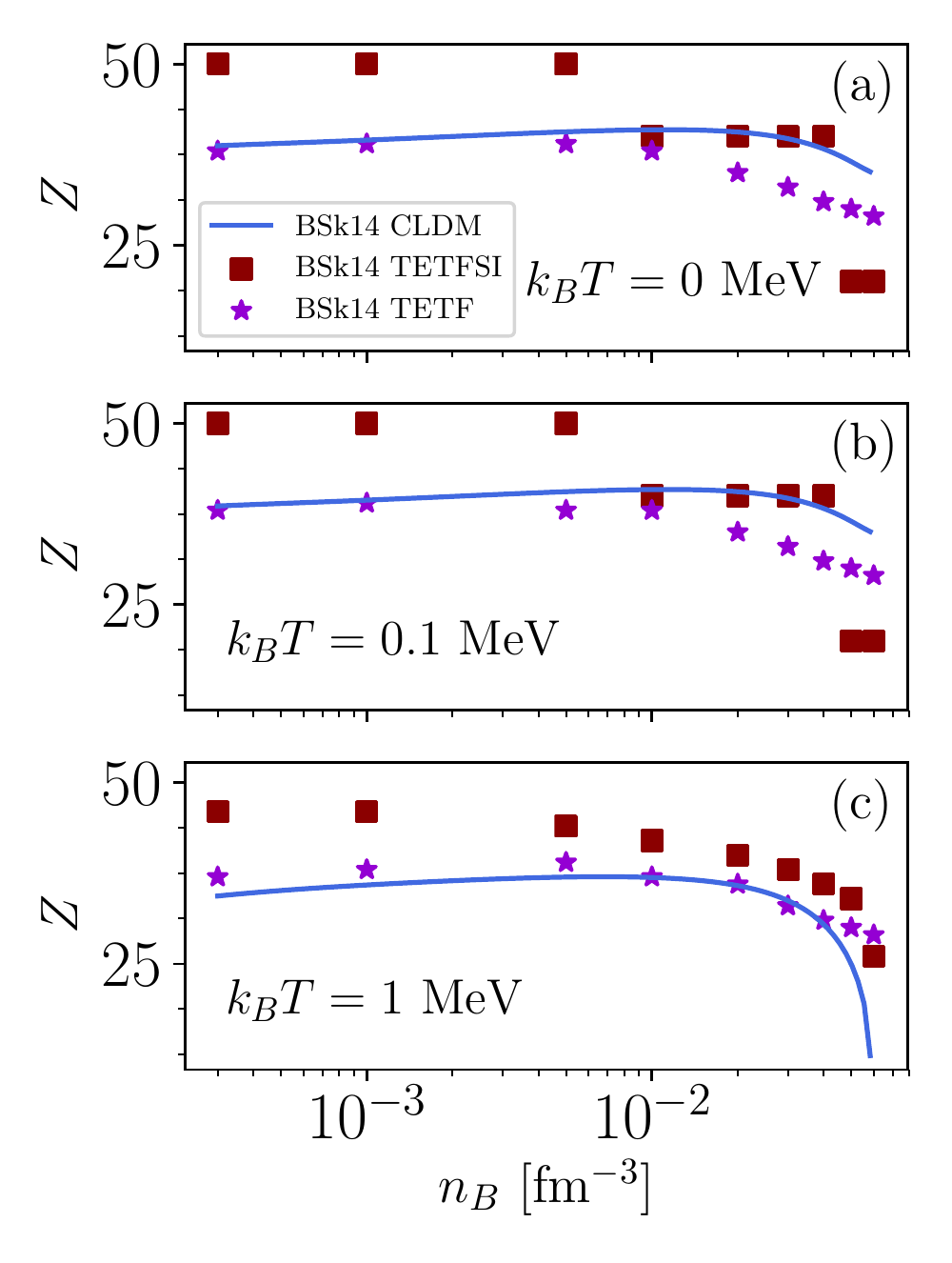}
\end{center}
\caption{Equilibrium value of proton number $Z$ as a function of baryon density
in the inner crust, for the BSk14 {functional}, for different temperatures: $k_B
T=0$~MeV (panel (a)), $k_B T=0.1$~MeV (panel (b)), and $k_B T=1$~MeV (panel (c)). Solid lines correspond to the {CLD model} calculations without shell effects, while square (star) symbols correspond to the {TETFSI (TETF)} results from~\cite{onsi2008}. See text for details.}
\label{fig:compo_bsk14}
\end{figure}

A study of the temperature dependence of shell effects {within the finite-Temperature Extended Thomas-Fermi plus Strutinsky Integral approach (TETFSI)} 
{using the BSk14 functional}
 was performed in 
\citet{onsi2008}, where it was shown that proton shell corrections decrease
substantially around $k_B T=1$~MeV (see their Tables~III-V).
This is shown in Fig.~\ref{fig:compo_bsk14}, where we compare, for three different temperatures, the proton value $Z$ in the inner crust as obtained in our CLD model without shell effects (solid line), with the 
results of~\cite{onsi2008} with (TETFSI, squares) and without (TETF, stars) shell effects.


{At relatively high temperature, our CLD results 
{are} in excellent agreement with the more microscopic 
TETF results of ~\cite{onsi2008}, {for which shell corrections are also neglected}.} 
It is clearly seen that the discontinuous behavior in $Z$ persists until $k_B T \approx 1$~MeV when the TETFSI results follow the same smooth trend as the TETF and our CLD results.
{At low temperature and high density, larger values of $Z$ are obtained in
the CLD {calculations} with respect to the TETF {ones}, showing the limits of the {model}. 
{Indeed, a precision of the order of 10$^{-4}$ MeV~fm$^{-3}$} is necessary to define the correct minimum (see Fig.~\ref{fig:fws})}. 


In view of the observed temperature dependence of shell effects in the microscopic calculations, we introduce a temperature-dependent factor to the zero-temperature shell corrections in the free energy of our CLD model,
\begin{equation}
\label{eq:shell-T}
F_{\rm nuc} = F_{\rm nuc}^{\rm CLD} + E_{\rm sh}(Z,T=0) x(T) \ ,
\end{equation}
where
\begin{equation}
\label{eq:shell-x}
x(T) \equiv \left( 1-\frac{2}{\pi} {\rm arctan} (\lambda T) \right) \ .
\end{equation}
The {coefficient} 
 $\lambda$ can thus be determined by two parameters, $T_0$ and $x(T_0)$, such that
\begin{equation}
\label{eq:shell-lambda}
\lambda = \frac{1}{T_0} \tan \left( \frac{\pi}{2} (1-x) \right) \ .
\end{equation}
We fix $T_0$, which represents the temperature at which shell effects vanish,
and $x(T_0)$ to reproduce the TETFSI results of \cite{onsi2008} (see also
Fig.~\ref{fig:compo_bsk14}), yielding $k_B T_0=1$~MeV and $x(T_0) = 0.02$.

This is of course a very rough treatment of the temperature dependence of shell effects, but we will show in the next section that the difference in the results obtained using or not this temperature dependence is smaller than the uncertainty 
{due to} our imperfect knowledge of the smooth part of the nuclear functional.

\section{Numerical results}
\label{sect:results}

\begin{figure}[htbp]
\begin{center}
\includegraphics[width=\linewidth]{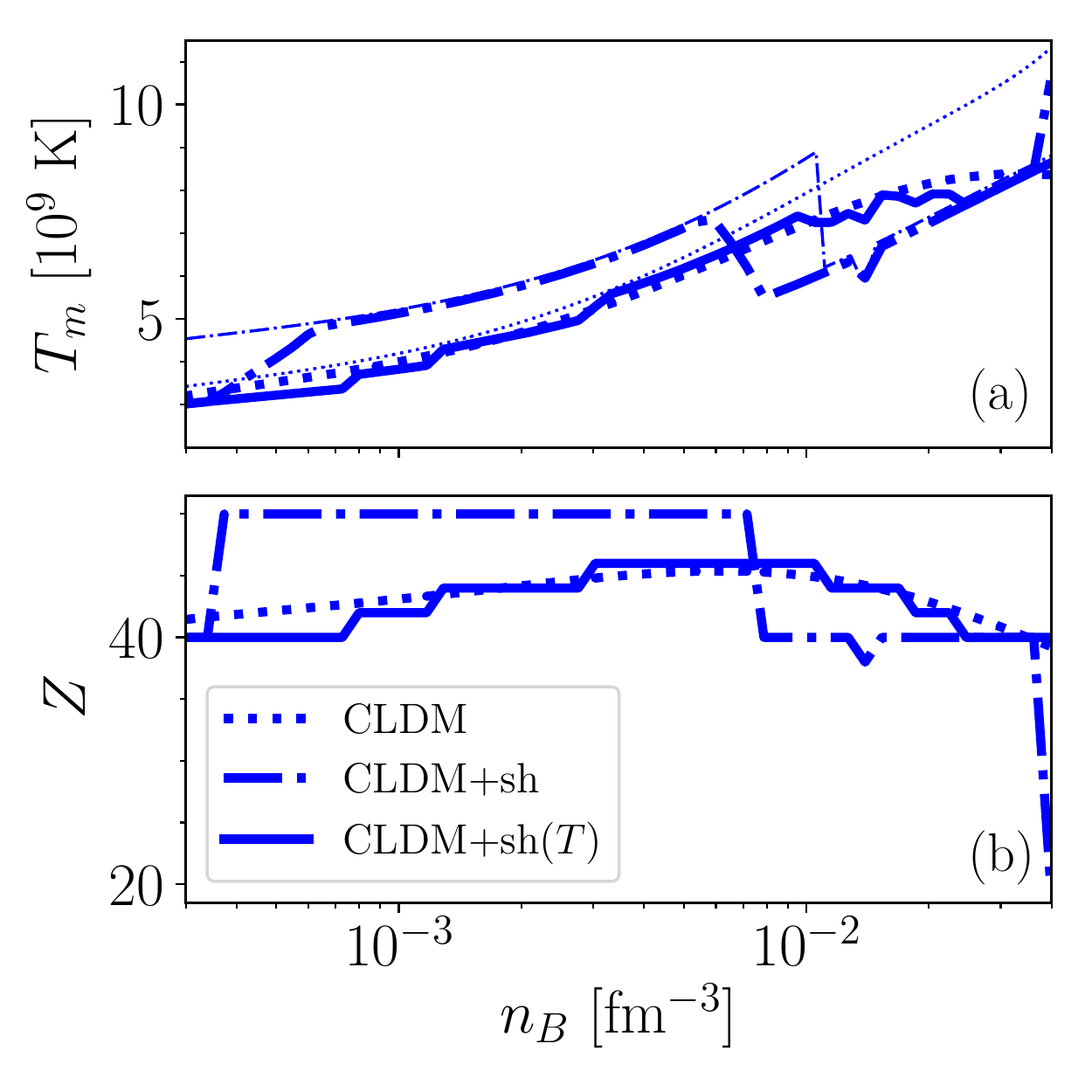}
\end{center}
\caption{Crystallization temperature $T_{\rm m}$ (panel (a)) and equilibrium value of $Z$ at $T=T_{\rm m}$ (panel (b)) as a function of baryon density in the inner crust, for the BSk25 {CLD {model}}. Dash-dotted (solid) lines correspond to the calculations including temperature-independent (temperature-dependent) shell effects. Results without shell effects are shown as dotted lines. Thick lines give self-consistent results for the crystallization temperature, while thin lines in the panel (a) correspond to
the approximate expression Eq.~(\ref{eq:Tm_book}). See text for details.}
\label{fig:tcryst25}
\end{figure}

For each BSk functional, the density domain ranging from the onset {of} neutron drip to the transition to homogeneous matter is explored. At each density, we progressively decrease the temperature from a high value {such that matter is in a liquid state, }
until the inequality $\mathcal{F}_{\rm liq}(T_{\rm m}) \ge \mathcal{F}_{\rm sol}(T_{\rm m})$ is verified. 
The corresponding value of $T_{\rm m}$ thus yields the crystallization temperature at that density.

\begin{figure}[htbp]
\begin{center}
\includegraphics[width=\linewidth]{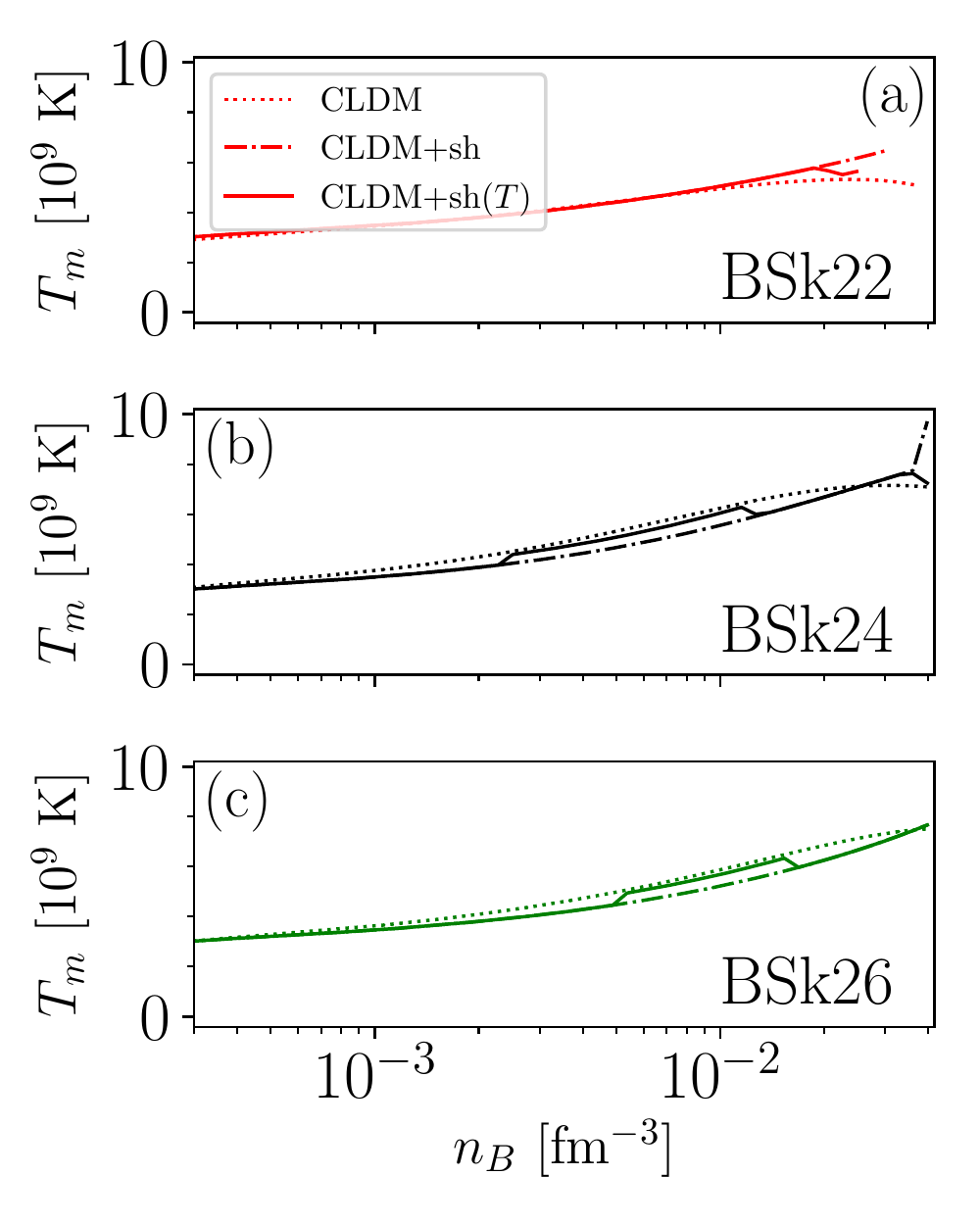}
\end{center}
\caption{Crystallization temperature $T_m$ as a function of baryon density in the inner crust, for the BSk22 (panel (a)), BSk24 (panel (b)), and BSk26 (panel (c)) {CLD {models}}. In each panel, dash-dotted (solid) lines correspond to the calculations including temperature-independent (temperature-dependent) shell effects. Results without shell effects are also shown (dotted lines). See text for details.}\label{fig:tcryst}
\end{figure}

The behavior of the crystallization temperature as a function of the baryon density is displayed in Fig.~\ref{fig:tcryst25} for the model BSk25, in comparison with the estimation from \citet{hpy2007} obtained in the limit of a pure Coulomb plasma,

\begin{equation}
  T_{\rm m} \approx \frac{(Ze)^2}{k_B a_N\Gamma_{\rm m}} \quad \text{K} \, . \label{eq:Tm_book}
\end{equation}
In this equation, 
the mean ion-coupling parameter at the crystallization point is defined as
$\Gamma_{\rm m}= (Ze)^{2}/(k_B a_N T_{\rm m}) \approx 175$ for a classical OCP,
and $a_N=\left(4\pi/(3V)\right)^{-1/3}$ with $V = A(1-n_g/n_0)/(n_B-n_g)$.

The qualitative behavior of the CLD calculation is very similar to the one obtained in \citet{hpy2007} applying Eq.~(\ref{eq:Tm_book}) to the ground-state composition obtained in the \citet{Douchin2001} CLD  model. Conversely, the calculation supposing temperature-independent shell effects exhibits discontinuities similar to the ones obtained {considering} the ground-state composition of the microscopic Hartree-Fock calculation of \citet{Negele1973} (see Fig.~3.17 of \citet{hpy2007}). These discontinuities are however considerably smoothed out if the expected modification of the shell structure with temperature is taken into account (full line in Fig. \ref{fig:tcryst25}).
We can see that the simple expression Eq.~(\ref{eq:Tm_book}) gives a fairly good estimation of the crystallization temperature
except at the highest densities. 
\cite{hpy2007} argued that setting $\Gamma_{\rm m} = 175$ is not reliable in the densest region of the crust because the amplitude of zero-point quantum vibrations of ions (treated as a small correction) becomes very large and even comparable to the lattice spacing. 


\begin{figure}[htbp]
\begin{center}
\includegraphics[width=\linewidth]{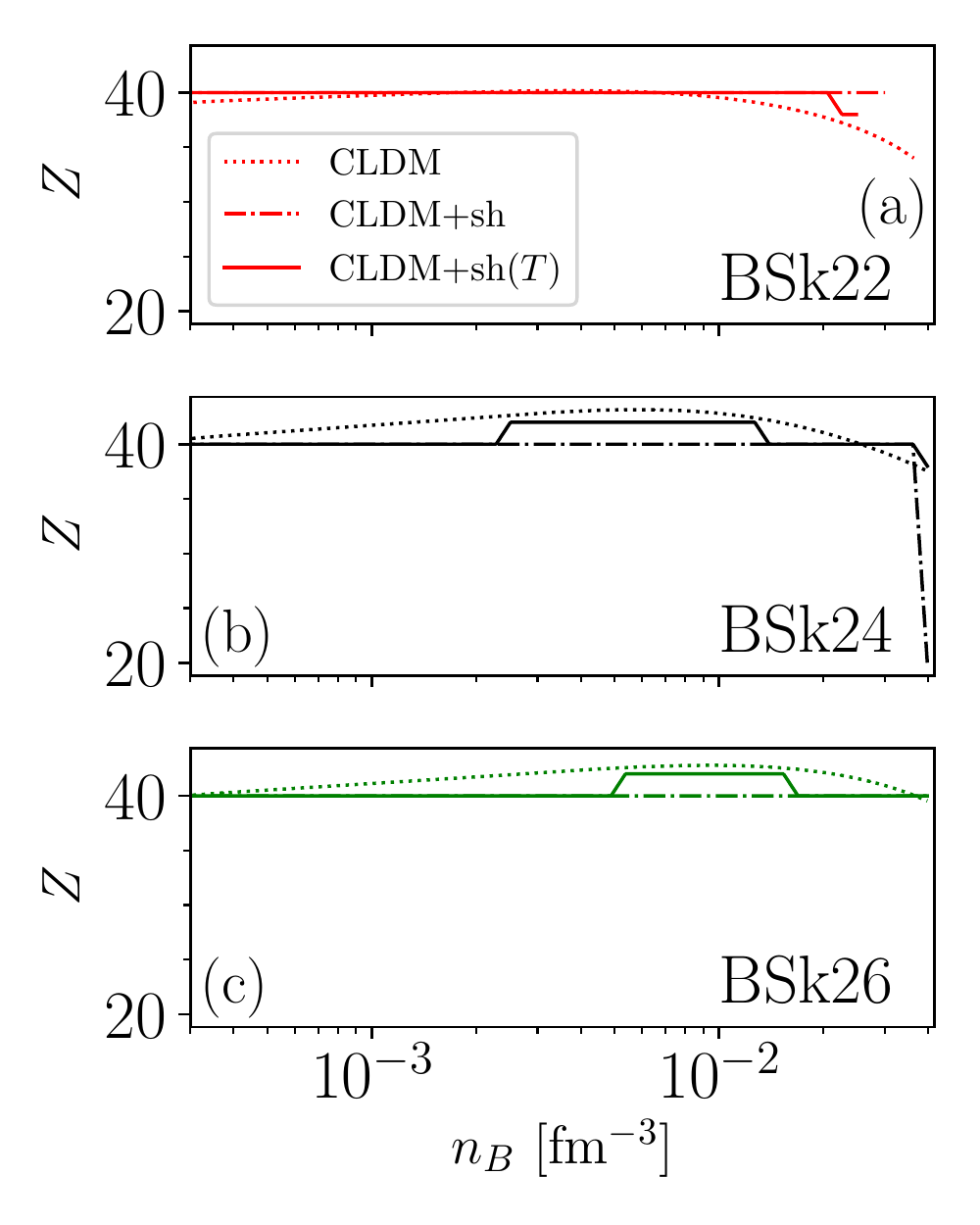}
\end{center}
\caption{Equilibrium value of proton number $Z$ as a function of baryon density in the inner crust at the crystallization temperature, for the BSk22 (panel (a)), BSk24 (panel (b)), and BSk26 (panel (c)) {CLD} models. In each panel, dash-dotted (solid) lines correspond to the calculations including temperature-independent (temperature-dependent) shell effects. Results without shell effects are also shown (dotted lines). 
See text for details.}\label{fig:compo_tcryst}
\end{figure}

The dependence on the nuclear model employed is further explored in Figs. \ref{fig:tcryst} and \ref{fig:compo_tcryst}, which show respectively the transition temperature and equilibrium value of the ion atomic number for the three different models  BSk22, BSk24, and BSk26.
The same qualitative trends as in Fig.~\ref{fig:tcryst25} can be observed. In particular, the discontinuous behavior of the crust composition obtained when the zero temperature composition is assumed, is considerably smoothed out if a temperature dependence of the shell effects is {introduced},  
with a global result very close to the CLD prediction. 
However, the different nuclear models lead to predictions which progressively diverge with {increasing depth},  
the crystallization temperature 
{differing by}
up to {$35 \%$} at the highest density close to the crust-core transition.

This model dependence arises from the lack of experimental {and} theoretical constraints for very neutron-rich nuclei and nuclear matter. As we can see from Eq.~(\ref{eq:fnuc}), this concerns both bulk properties, that is the nuclear EoS of asymmetric matter, and surface properties, that is the surface tension of extremely neutron-rich nuclei. 
To progress on this issue, it is important to know the relative weight of the bulk and surface properties in the determination of the uncertainties in the inner-crust {properties}. 

\begin{figure}
\begin{center}
\includegraphics[width=\linewidth]{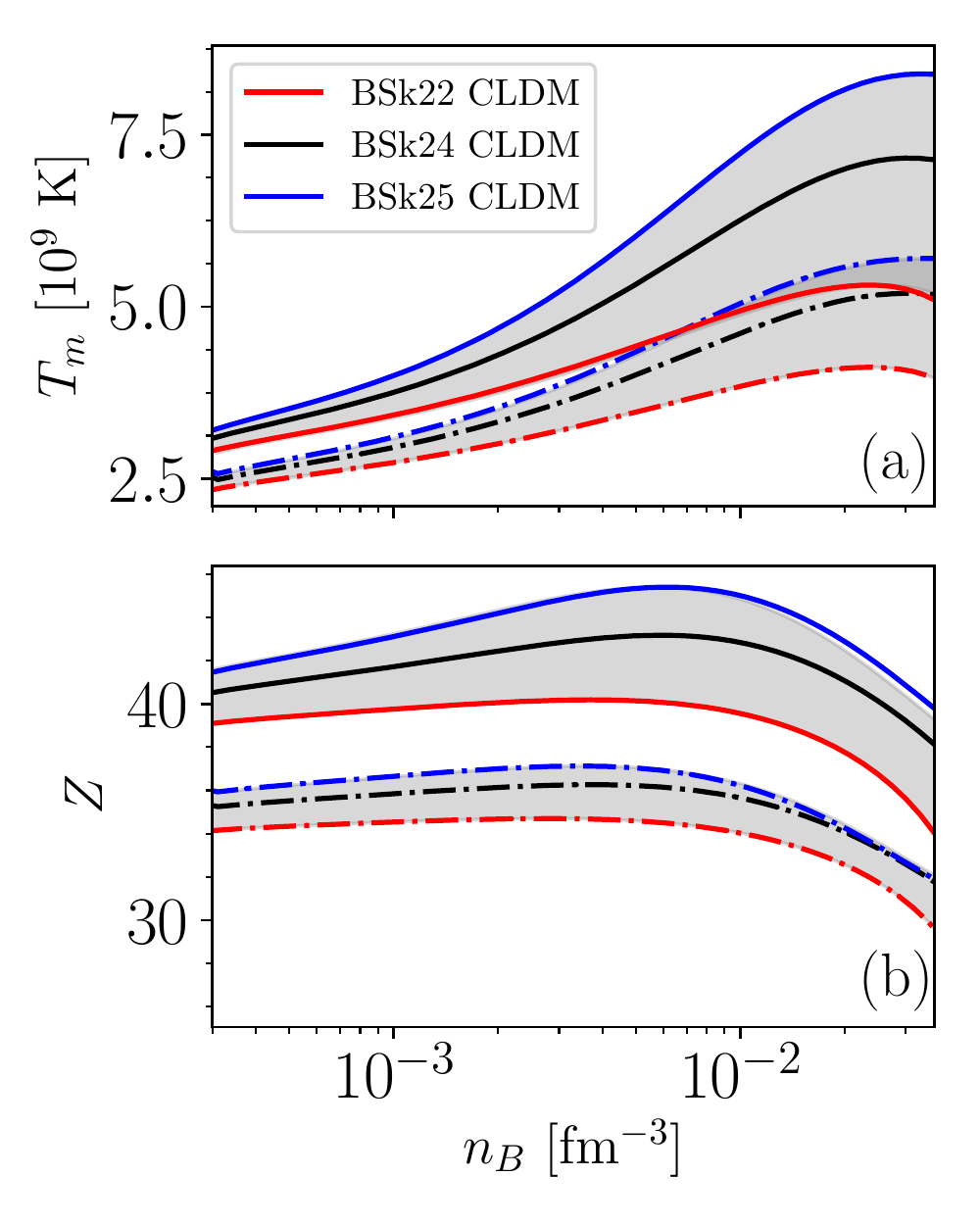}
\end{center}
\caption{{Crystallization temperature $T_{\rm m}$ (panel (a)) and equilibrium value of $Z$ (panel (b)) as a function of baryon density
in the inner crust for BSk22, BSk24, and BSk25 CLD {models} with surface parameters 
fitted to spherical nuclei (dot-dashed lines), or to associated ETF 
calculations (solid lines). Grey bands represent extrema. 
See text for details.}}
\label{fig:tm_compo_final1}
\end{figure}

\begin{figure}
\begin{center}
\includegraphics[width=\linewidth]{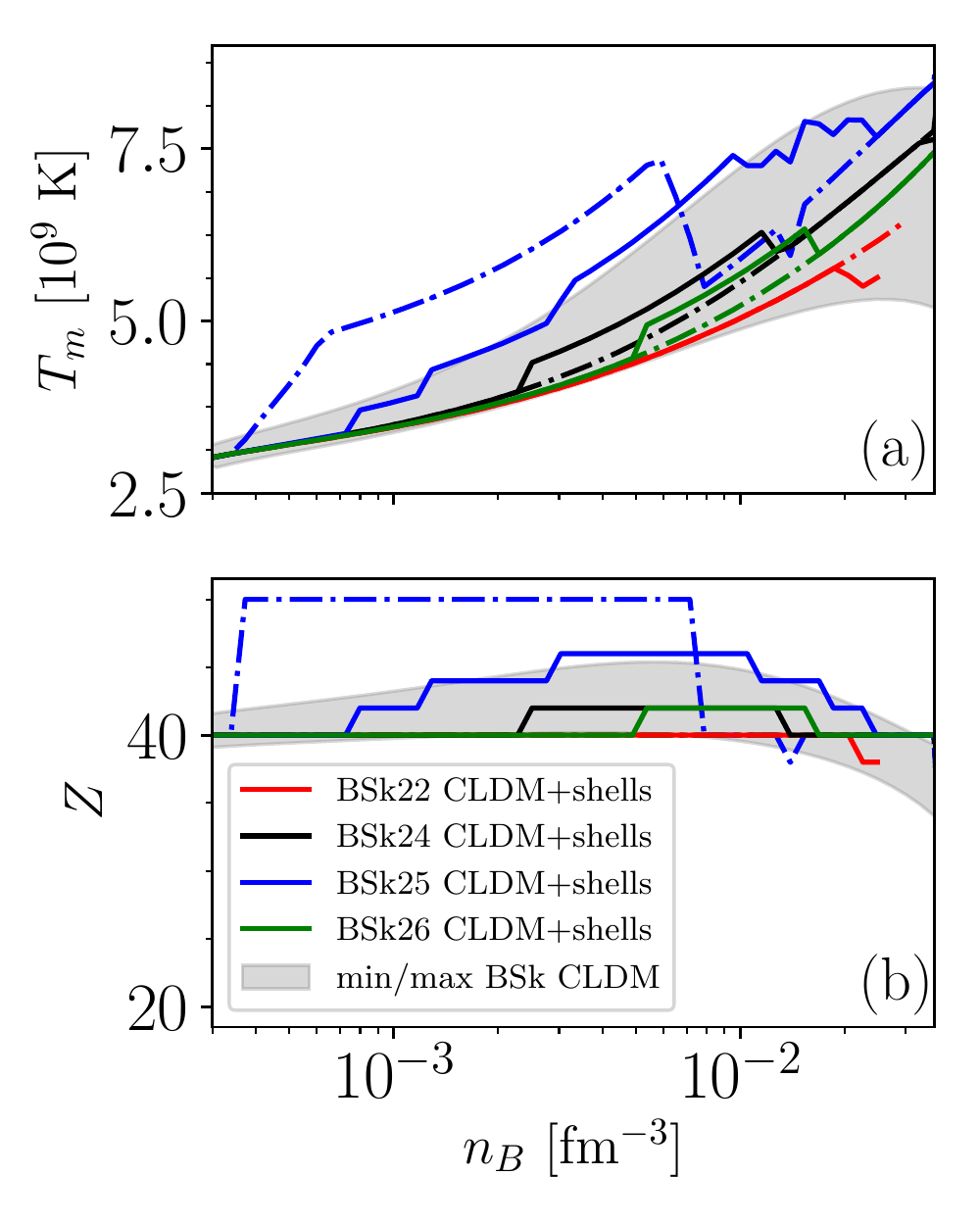}
\end{center}
\caption{{Crystallization temperature $T_{\rm m}$ (panel (a)) and equilibrium 
    value of $Z$ (panel (b)) as a function of baryon density
in the inner crust. The grey band corresponds to
extrema for the {BSk22, BSk24, and BSk25} CLD {models} with surface parameters fitted to associated ETF 
calculations. Solid (dot-dashed) lines represent the calculations for
BSk22, BSk24, BSk25, and BSk26 {CLD {models}} with temperature-dependent
(temperature-independent) shell corrections. See text for details.}}
\label{fig:tm_compo_final2}
\end{figure}

To this aim, we have repeated the CLD calculations with the four different EoSs, and we have fixed the surface properties following \citet{Carreau2019}, {by} 
fitting only the experimentally measured masses of the spherical magic and semi-magic nuclei $^{40,48}$Ca, $^{48,58}$Ni, $^{88}$Sr, $^{90}$Zr, $^{114,132}$Sn, and $^{208}$Pb. 
 The $p$ parameter governing the behavior of the surface tension Eq.~(\ref{eq:sigma}) at extreme isospin values, is completely unconstrained by this fit, and it is fixed 
to the constant value $p=3$ suggested in the seminal work by ~\citet{Ravenhall1983}. 
The resulting models then differ for their bulk properties, but correspond to comparable surface properties consistent with  experimental {data}, 
but unconstrained in the extreme neutron-rich regime. 

{The predictions for the crystallization temperature with {the} BSk22, BSk24, and BSk25 
  {CLD} models are shown in panel (a) of Fig. \ref{fig:tm_compo_final1} (dotted lines), while
the associated equilibrium value of $Z$ is reported in panel (b).} The width of the bands can be interpreted as an estimate of the uncertainty on the
respective observables, due to our incomplete knowledge of the nuclear EoS.
{In the same figure, solid lines represent the CLD results already shown in Figs.~\ref{fig:tcryst25},~\ref{fig:tcryst},~\ref{fig:compo_tcryst}, with surface parameters optimized on ETF calculations performed up to the respective driplines. 
  A systematic difference can be observed between the two bands: systematically higher $Z$ values are obtained when surface tension of neutron-rich nuclei is optimized on 
the microscopic models, and this difference obviously reflects on the crystallization temperature, which roughly scales a $Z^{5/3}$ (see Eq.~(\ref{eq:Tm_book})).}

{It is important to stress that the bulk parameters of the BSk functionals were precisely fitted to both properties {of finite nuclei} and ab-initio neutron-matter calculations, see~\cite{goriely2013}. 
	The residual uncertainty  in the nuclear EoS is not negligible, but its consequence is less important than the uncertainties on  the surface energy for neutron-rich nuclei close to neutron drip. The latter can then be considered as the key physical quantity determining the crust composition and crystallization temperature.} 
{It is also interesting to observe the anti-correlation between the symmetry 
energy {coefficients} and the crystallization temperature in 
the inner crust. Indeed, we can see that in this regime, the {higher the symmetry energy at saturation}, 
the lower the crystallization temperature, with $E_{sym} = 32$ MeV, 
$E_{sym} = 30$ MeV, and $E_{sym} = 29$ MeV, for BSk22, BSk24, and BSk25, 
respectively.}

{In Fig.~\ref{fig:tm_compo_final2}, we report the band for the CLD 
  results with surface parameters optimized on ETF calculations and we show the 
  results with temperature-independent (temperature-dependent) shell 
  corrections in dot-dashed (solid) lines. Apart {from} BSk25 under the extreme and quite {unrealistic} hypothesis 
that shell effects are not affected by temperature, it is seen that all 
calculations with shell effects fit in the CLD band. This indicates that simple CLD 
modeling could be sufficient to study crust properties at crystallization.}

\begin{figure}
\begin{center}
\includegraphics[width=\linewidth]{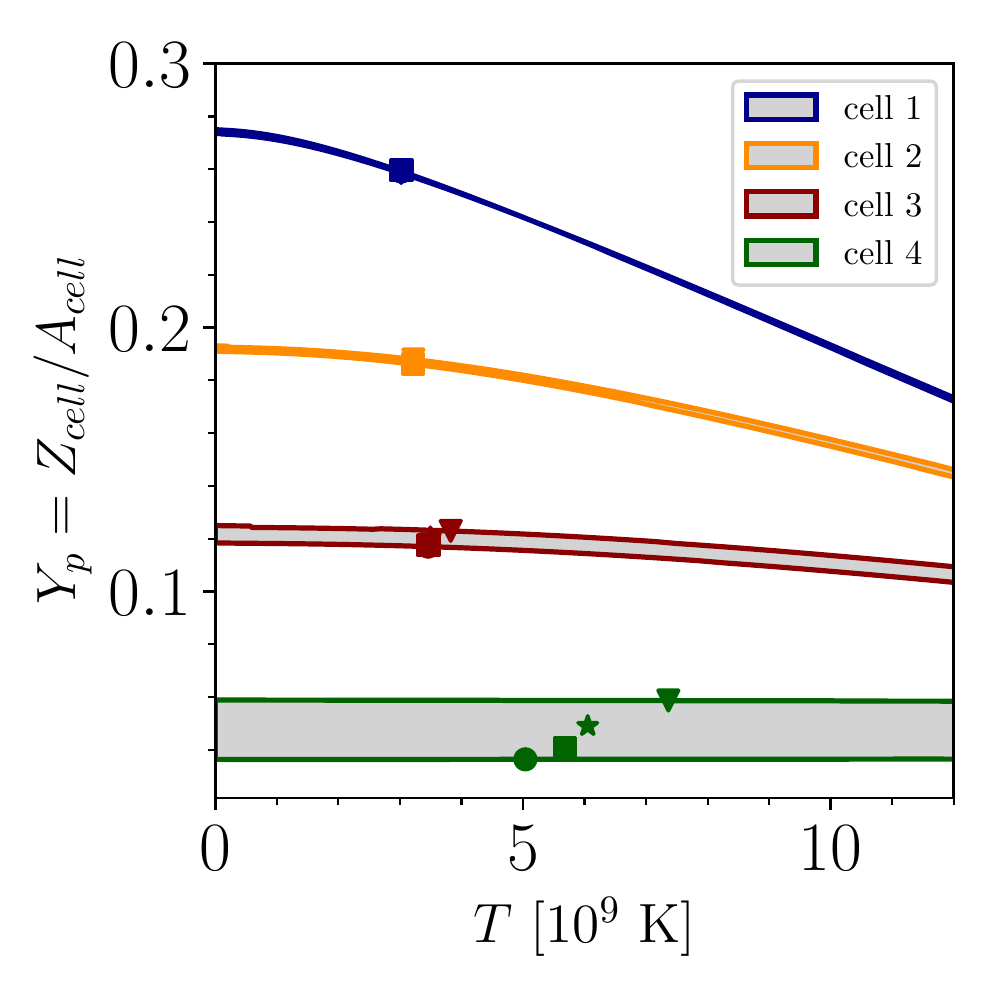}
\end{center}
\caption{{Proton fraction $Y_p$ as a function of 
temperature for different values of the baryon density: $n_B=3\times 10^{-4}$ \fm~
(cell 1), $n_B = 5\times 10^{-4}$ \fm~(cell 2), $n_B = 10^{-3}$ \fm~(cell 3), 
and $n_B = 10^{-2}$ \fm~(cell 4). Grey bands represent minima and maxima of {CLD
calculations for the different BSk functionals including temperature-dependent shell 
corrections}. Circles, stars, triangles, and squares correspond to crystallization 
temperature for BSk22, BSk24, BSk25, and BSk26, respectively. See text for details.}}
\label{fig:ypcell}
\end{figure}

{Finally, the variation of the {proton fraction $Y_p$ (or, equivalently, of the electron fraction $Y_e=Y_p$)} with temperature in the 
  inner crust is shown in Fig.~\ref{fig:ypcell}. 
Four different densities are considered. We can see that the equilibrium proton fraction at crystallization 
(marked by the solid symbols) {tends} to be lower than the prediction for fully catalyzed matter at $T=0$. 
The effect is {negligible in the densest} part of the inner crust close to the crust-core transition, 
where the catalyzed matter hypothesis appears well justified. 
Conversely, a sizeable difference is observed in the proximity of the outer crust close to the drip point. 
In Fig.~\ref{fig:ypcell}, the four BSk CLD models are considered and, for each density, 
the dispersion in their predictions {is represented} by the grey bands. These bands become thinner than 
the width of the curves for the lowest densities, meaning that the results are independent of the model.

{The decrease of the proton fraction with the temperature is easily understood from the weak interaction
 equilibrium condition $\mu_e=\mu_n-\mu_p$: with increasing temperature, the electron chemical potential 
$\mu_e$  increases leading to an increasing difference between the neutron and proton chemical potential 
$\mu_n$ and $\mu_p$, and therefore an increasing neutron-proton asymmetry. 
}
Because 
of the typical time scale of the weak interaction relative to the cooling dynamics, 
it is possible that the composition might be frozen at a 
temperature even higher than the crystallization temperature. In that case,  the catalyzed matter hypothesis 
would not be justified, and the electron fraction 
estimated at $T=0$ might lead to a {significant} overestimation of the effective electron fraction of the crust;
this conclusion does not depend on the adopted nuclear model, and the effect  is expected to be amplified in the outer 
crust.}

\section{Conclusions}
\label{sect:conclus}

In this work we have presented an extension of the recent work of
\citet{fantina2019}, addressing the crystallization temperature and associated
composition of the inner crust of a non-accreting neutron star. In particular
we have challenged the cold catalyzed-matter hypothesis by performing
consistent calculations in the liquid and solid phases to determine the
crystallization temperature of the crust in the OCP approximation, and the
possible modifications with respect to the ground-state composition. To settle
the model dependence of the results, four different up-to-date nuclear
functionals were considered. The crystallization temperature was shown to be
systematically lower than the analytic expectation from \citet{hpy2007} based
on a pure Coulomb plasma. Despite this fact, sufficiently high values are
obtained in the innermost part {suggesting} that substantial reduction of the
shell effects should take place, and the simple CLD approximation should lead
to reasonably good results. The actual importance of shell effects is affected
by considerable uncertainties, and a very crude approximation was employed in
this work. Still, the highest source of model dependence comes from the smooth
part of the nuclear functional. In particular, a full handle on the surface
properties at extreme isospin values is seen to be essential to fix the
composition of the crust at crystallization and the associated temperature. 

Our semi-classical formalism based on the CLD model is only an approximation of the more complete TETFSI calculations that have been performed in ~\citet{onsi2008}. The main advantage of this formalism is that the computational time is significantly reduced in comparison with that of TETFSI calculations, in particular if one wants to evaluate the crystallization temperature. Also, this formalism can be easily extended to account for impurities, along the same line as we have done in~\citet{fantina2019} for the outer crust. The extension to a multicomponent plasma approach and the computation of the impurity factor is in progress, and will be presented in a forthcoming publication.


\begin{acknowledgements}
This work has been partially supported by the IN2P3 Master Project MAC, the CNRS PICS07889, and the PHAROS European Cooperation in Science and Technology (COST) action CA16214.
The work of N.C. was supported by Fonds de la Recherche Scientifique (Belgium) under grant IISN 4.4502.19.
J.M.P. thanks ULB for the award of a "Chaire internationale" for the month of November, 2019.
\end{acknowledgements}



\end{document}